\documentclass[12pt]{article}\usepackage{AllMomsSimplex}

\begin{document}

\title{All Moments of the Uniform Ensemble\\
of Quantum Density Matrices}

\author{Robert R. Tucci\\
        P.O. Box 226\\
        Bedford,  MA   01730\\       tucci@ar-tiste.com}

\date{ \today}

\maketitle

\vskip2cm
\section*{Abstract}
Given a uniform ensemble of quantum density matrices $\rho$,
it is useful to calculate the mean value over this ensemble of
a product of entries of $\rho$. We show how to calculate such moments
in this paper. The answer involves well known results from
Group Representation Theory and Random 
Matrix Theory.
This quantum problem has 
a well known classical counterpart:
given a uniform ensemble of probability distributions 
$P=(P_1, P_2, \ldots, P_N)$
where the $P_j$ are non-negative reals that sum to one,
calculate the mean value over this probability simplex
of products of $P$ components. The answer to the classical
problem follows from an integral formula due to Dirichlet.

\newpage
\section{Introduction}

The probability simplex of dimension $N-1$ 
is the set of all points $P=(P_1, P_2, \ldots, P_N)$,
where the components of $P$ are
non-negative reals that sum to one.
Take $N=3$ for simplicity.
It is useful to calculate moment integrals such as:

\beq
\int d^3P\;\;P_1^2 P_3
\;,
\eeq
where $d^3P = dP_1dP_2 dP_3$, 
and where the point $P$ ranges over the 2 dimensional probability simplex.
As will be discussed in detail later,
such integrals can be performed using
an integral formula due to Dirichlet\cite{Jef}\cite{Whit}.

Now suppose we generalize this problem to the 
quantum regime by considering quantum density matrices
instead of classical probability distributions.
Suppose that $\rho$ is a quantum density matrix 
(i.e., $\rho$ is a Hermitian matrix with non-negative eigenvalues that
sum to one). It is useful to calculate moment integrals such as

\beq
\int \fatd\rho \;\; (\rho_{1,1})^2 \rho_{1,2}
\;.
\label{eq:quant-int-eg}\eeq
As will be discussed in detail later,
there is a very natural way of defining
the measure $\fatd\rho$. The goal of this paper is 
to show how to calculate integrals like Eq.(\ref{eq:quant-int-eg}).
The answer involves well known results from
Group Representation Theory
\cite{FulHar}
\cite{Ham}
\cite{Sch}
\cite{Weyl},
and Random Matrix Theory
\cite{MeBook}
\cite{ItZu80}
\cite{Me81}.
Ref.\cite{ItZu80}
by Itzykson and Zuber is especially pertinent
to this paper.

\section{Notation}
In this section, we will introduce some notation that
will be used in subsequent sections.

RHS (ditto, LHS) will mean ``right hand side" (ditto, ``left hand side").
For any complex number $z$,
let $z_\Re$ (ditto, $z_\Im$) represent its real (ditto, imaginary) part.

As usual, $\delta(x)$ for real $x$ will denote the Dirac delta function  
$\delta(x) = \int_{-\infty}^\infty \frac{dk}{2\pi}e^{ikx}$.
Likewise, $\delta(x, y)$ and $\delta_x^y$ will denote the Kronecker
delta function. $\delta_x^y$ equals 1 if $x=y$ and it equals 0 if $x\neq y $.
$\epsilon_{i_1, i_2, \ldots, i_N}$ will denote the totally 
anti-symmetric tensor with $N$ indices. $\epsilon_{i_1 ,i_2, \ldots, i_N}$
equals 1 if
$(i_1, i_2, \ldots, i_N)$ is
an even permutation of $(1,2,3, \ldots, N)$, and 
it equals $-1$ for odd
permutations. Let $\theta(\calS)$ be the ``truth function"
or ``indicator function"; it equals 1 
if the statement $\calS$ is true, and it equals 0 if $\calS$ is false. 
For example,
$\theta(x> 0)$ is the unit-step function; it equals 1 if $x>0$ and
it equals 0 if $x\leq 0$.

The set of Hermitian $N\times N$ matrices will 
be denoted by $Herm(N)$.
Any square matrix 
$A$ is said to be positive semi-definite (ditto, positive definite)
if the eigenvalues of $A$ are non-negative (ditto, strictly positive).
We will write $A\geq 0$ 
 (ditto, $A> 0$ )
if $A$ is positive semi-definite (ditto, positive definite). 
Suppose 
$A$ and $B$ are two $N\times N$ matrices.
We define the dot product of $A$ and $B$ by
$A\cdot B = \tr(A^T B) = \sum_{i,j} A_{i,j}B_{i,j}$.
We will use $[\pder{}{A_{i,j}}]_{i, j} = \pder{}{A}$
to denote the matrix of the partial derivatives with respect to 
the entries of $A$.
Note that
$B\cdot \pder{}{A} = \tr(B^T\pder{}{A})$

We will often denote the set $\{x_k : k\in K\}$ 
by $\{x_k\}_{k\in K}$, 
or simply by $\{x_k\}_k$
when the set $K$ is clear from the context.
Likewise, the ordered set 
$(x_k : k\in K)$ 
will be denoted by $(x_k)_{k\in K}$ 
or simply by $(x_k)_k$.
$\sum_{k\in K} x_k$ 
will often be denoted by $\sum \{x_k\}_k$
and 
$\prod_{k\in K} x_k$ 
by $\prod \{x_k\}_k$.

Suppose $\alpha$ is an $N$ component column vector,
$\alpha = (\alpha_0, \alpha_1, \ldots, \alpha_{N-1})^T$.
One defines the {\it Vandermonde determinant} of $\alpha$
by
\beq
\Delta(\alpha) = 
= 
\left |
\begin{array}{ccccc}
1 & \alpha^1_0 & \alpha^2_0 &\ldots &\alpha^{N-1}_0 \\
1 & \alpha^1_1 & \alpha^2_1 &\ldots &\alpha^{N-1}_1 \\
\vdots & \vdots & \vdots &  &\vdots\\
1 & \alpha^1_{N-1}& \alpha^2_{N-1} &\ldots &\alpha^{N-1}_{N-1}
\end{array}
\right|
\;.\eeq
Often, we will express the RHS  of the last equation 
more succinctly as follows:

\beq
\Delta(\alpha) = \det(\alpha^0, \alpha^1, \ldots, \alpha^{N-1})
\;.
\eeq
As is well known and easily proven:

\beq
\Delta(\alpha) = \prodset{\alpha_j - \alpha_i}{0\leq i < j\leq N-1}
\;.
\label{eq:diff-prod}\eeq 
Vandermonde determinants are sometimes, especially in very old literature, 
referred to as {\it difference products},
or as basic {\it  alternants}.

We end this section by  defining symbols for two 
expressions that arise frequently in subsequent sections.
Let

\beq
\calF_N = \prodset{j\;!}{1\leq j \leq N}
\;.
\eeq
From Eq.(\ref{eq:diff-prod}), it follows that
$\calF_N=\Delta(0,1,2, \dots, N)$.
Let

\beq
L_N = \frac{N(N-1)}{2}
\;.
\eeq
An $N\times N$ matrix
has $L_N$ entries below (and above)  its main diagonal.
Also, $L_N = 1 + 2 +\cdots+ (N-2) + (N-1)$.

\section{Ensemble of Classical Probability Distributions}
In this section, we will review the simplex moments integral and its variants.

Suppose index $b$ ranges over the integers from 1 to $N_\rvb$.
Define the following moment integral over 
an $N_\rvb-1$ dimensional simplex:

\begin{subequations}
\label{eq:sim-int}
\beq
\calS_{N_\rvb-1}\equiv
\prodset { \int_0^\infty dx_b }{b}
\delta (\sum_b x_b - \lambda)
\prodset { x^{\nu_b}_b }{b}
\;, 
\label{eq:sim-int-a}\eeq
where $\lambda$ is a positive real and the
$\nu_b$ are non-negative integers.
It is well known that 

\beq
 \calS_{N_\rvb-1}=
\frac{
\prodset{ \nu_b!}{b}
\lambda^\nu 
}{ \nu ! }
\;, 
\label{eq:sim-int-b}
\eeq
where

\beq
\nu = \sum_b \nu_b + N_\rvb -1
\;. 
\label{eq:sim-int-nu}
\eeq\end{subequations}
Eqs.(\ref{eq:sim-int})  can be
generalized so that they also apply  to non-integer
$\nu_b$ (just replace factorials by Gamma functions
according to the prescription $n!\rarrow\Gamma(n+1)$), 
but we won't bother with such generalizations in this paper.
Eqs.(\ref{eq:sim-int}) are a generalization to higher dimensions
of the {\it Beta function}, given by 
\beq
\beta(m, n) = \int^1_0 dt \;t^{m-1} (1-t)^{n-1}
= \frac{\Gamma(m) \Gamma(n)}{\Gamma(m+n)}
\;.
\eeq

Now let $B$ be an index that ranges over the integers from 1 to $N_\rvB$, 
where $N_\rvB = N_\rvb-1$. The $N_\rvB$
dimensional {\it Dirichlet  integral} is defined by 

\begin{subequations}\label{eq:dir-int}
\beq
{\cal D}_{N_\rvb-1} \equiv
\prodset { \int_0^\infty dx_B }{B}
\theta(\sum_B x_B < \lambda)
\prodset { x^{\nu_B}_B }{B}
f(\sum_B x_B)
\;. \label{eq:dir-int-a}
\eeq
It is well known that 

\beq
 {\cal D}_{N_\rvb-1}=
\frac{
\prodset{ \nu_B!}{B}
}{ \nu ! }
g(\lambda)
\;, \label{eq:dir-int-b}
\eeq
where

\beq
g(\lambda) =
\int_0^\lambda dt\;f(t) t^{\nu-1} \nu
\;.
\eeq \end{subequations}
Note that $g(\lambda) = \lambda^\nu$
when $f(t)=1$.
In Eqs.(\ref{eq:dir-int}), 
$\nu$ is defined as in Eq.(\ref{eq:sim-int-nu})
but with $\nu_{N_\rvb}=0$ so that 
$\sum_b \nu_b = \sum_B \nu_B$.

Proofs of Eqs.(\ref{eq:sim-int}) for $\calS_{N_\rvB}$
and Eqs.(\ref{eq:dir-int}) for $\fatd_{N_\rvB}$
are easily found in the literature (see, for example Refs.
\cite{Jef}\cite{Whit}) so we won't present them here.
However, we do want to emphasize that these
two integral formulas follow trivially from each other.
One can prove as follows that the formula for $\fatd_{N_\rvB}$
implies the formula for $\calS_{N_\rvB}$. Take $\frac{d}{d\lambda}$ of 
Eqs.(\ref{eq:dir-int-a}) and (\ref{eq:dir-int-b}). Then 
use $\frac{d}{dx} \theta(x>0) = \delta(x)$ and  the
fact that $N_\rvB = N_\rvb - 1$.
Conversely, one can prove as follows that the formula for $\calS_{N_\rvB}$
implies the formula for $\fatd_{N_\rvB}$. Applying the operator 
$\Omega =\int_0^{\lambda'} d\lambda\; \frac{d}{d\lambda}$
to Eq.(\ref{eq:sim-int-b}) gives

\beq
\int_0^{\lambda'} d\lambda\; f(\lambda) 
\frac{\calS_{N_\rvb -1}}{d\lambda} =
\frac{\prodset{\nu_b}{b}}{\nu!}
\left (\int_0^{\lambda'} d\lambda\; f(\lambda) \lambda^{\nu-1} \nu \right)
\;,
\eeq
whereas applying $\Omega$ to Eq.(\ref{eq:sim-int-a}) gives

\begin{subequations}\label{eq:sim-to-dir}
\beqa 
\lefteqn{
\int_0^{\lambda'} d\lambda\; f(\lambda) \frac{\calS_{N_\rvb -1}}{d\lambda} = 
} \\
\label{eq:sim-to-dir-b}
&&= \prodset{\int^\infty_0 x_b\; x^{\nu_b}_b}{b} 
\int_0^{\lambda'}d\lambda\; f(\lambda) 
\left( (-1) \pder{}{x_{N_\rvb}} \right)\delta(\sum_b x_b -\lambda)
\\
\label{eq:sim-to-dir-c}
&&=\prodset{\int^\infty_0 dx_B\; x^{\nu_B}_B}{B} 
\int_0^{\lambda'} d\lambda\; f(\lambda) \delta(\sum_B x_B - \lambda)
\\
&&= {\cal D}_{N_\rvb-1}
\;
\eeqa 
\end{subequations}
In going from line $b$ to
 line $c$, we
 have assumed that $\nu_{N_\rvb}=0$
and performed the $x_{N_\rvb}$ integration. 

\section{Some Results from \\
Group  Representation Theory} 
In this section, we will  review quickly some well known facts from 
Group Representation Theory.
For more details and proofs, see, for example, Refs.
\cite{FulHar}
\cite{Ham}
\cite{Sch}
\cite{Weyl}.

A $K$-box Young graph with at most $N$ rows is
specified by an $N$-dimensional column vector of integers
$\eta = (\eta_0, \eta_1, \ldots, \eta_{N-1})^T$
such that $\eta_0\geq\eta_1\geq\cdots\geq\eta_{N-1}\geq 0$
and $\sum_{j=0}^{N-1} \eta_j = K$.
$\eta_j$ is the number of 
boxes in the $j$-th row, where the top row is the 0-th one.
Each subsequent row has the same number or fewer
boxes than the row above.
If some of the last few components of $\eta$ are zero,
they are often omitted.
For example, 
$$
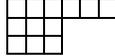
\tiny\yng(6,3,3)
$$
is specified by $\eta=(6,3,3,0, 0)^T$ with $K=12$ and $N=5$.

If $S_K$ is the symmetric group (permutation  group) on $K$ letters,
then the classes of $S_K$ 
are specified by 
a $K$-tuple of non-negative integers
$i =(i_1, i_2, \ldots, i_K)$ such that $1 i_1 + 2 i_2 +\ldots  +Ki_k= K$.
$\calC(i) = (1^{i_1}, 2^{i_2},\ldots, K^{i_K})$
represents the class of elements 
of $S_K$ with
$i_1$ cycles of length 1, 
$i_2$ cycles of length 2, \ldots, and 
$i_K$ cycles of length $K$.
The order (i.e., number of elements)  of $\calC(i)$ is given by

\beq
|\;\calC(i)| = 
\frac{K!}
{ (1^{i_1}i_1!)(2^{i_2}i_2!)\cdots (K^{i_K}i_K!)}
\;.
\eeq
There is a one-to-one onto correspondence between:
(1)the  irreps of $S_K$,
(2)the classes of $S_K$,
(3)the Young graphs with $K$ boxes.
Tables \ref{tab:perm-1} to \ref{tab:perm-4}
give the characters for each (irrep, class) 
pair of $S_K$, where $K$ ranges from 1 to 4.

\begin{table}[h]
\begin{center}
\begin{tabular}{c|c|c}
K=1& $(1)$ &{\small $\leftarrow$class}\\
{\small Characters}$\searrow$& 1&{\small $\leftarrow$order}\\
\hline{\tiny \yng(1)}& 1 &\\
\cline{1-2}
{\small $\uparrow$ irrep}\\
\end{tabular}
\caption{Characters for each (irrep, class) 
pair of the permutation group $S_1$.}
\label{tab:perm-1}
\end{center}
\end{table}

\begin{table}[h]
\begin{center}
\begin{tabular}{c|c|c|c}
K=2& $(1^2)$ & $(2)$&{\small $\leftarrow$class}\\
{\small Characters}$\searrow$& 1& 1&{\small $\leftarrow$order}\\
\hline{\tiny \yng(2)}& 1&1 &\\
\cline{1-3}
{\tiny \yng(1,1)}& 1&-1 &\\
\cline{1-3}
{\small $\uparrow$ irrep}\\
\end{tabular}
\caption{Characters for each (irrep, class) 
pair of the permutation group $S_2$.}
\label{tab:perm-2}
\end{center}
\end{table}

\begin{table}[h]
\begin{center}
\begin{tabular}{c|c|c|c|c}
K=3& $(1^3)$ & $(1,2)$& $(3)$&{\small $\leftarrow$class}\\
{\small Characters}$\searrow$& 1& 3&2&{\small $\leftarrow$order}\\
\hline{\tiny \yng(3)}& 1&1&1&\\
\cline{1-4}
{\tiny \yng(2,1)}&2 &0&-1 &\\
\cline{1-4}
{\tiny \yng(1,1,1)}&1 &-1&1 &\\
\cline{1-4}
{\small $\uparrow$ irrep}\\
\end{tabular}
\caption{Characters for each (irrep, class) 
pair of the permutation group $S_3$.}
\label{tab:perm-3}
\end{center}
\end{table}

\begin{table}[h]
\begin{center}
\begin{tabular}{c|c|c|c|c|c|c}
K=4& $(1^4)$ & $(1^2,2)$& $(1,3)$& $(2^2)$& $(4)$&{\small $\leftarrow$class}\\
{\small Characters}$\searrow$& 1& 6&8&3&6&{\small $\leftarrow$order}\\
\hline
{\tiny \yng(4)}& 1&1&1&1&1&\\
\cline{1-6}
{\tiny \yng(3,1)}&3 &1&0 &-1&-1&\\
\cline{1-6}
{\tiny \yng(2,2)}&2 &0&-1 &2&0&\\
\cline{1-6}
{\tiny \yng(2,1,1)}&3 &-1&0&-1&1&\\
\cline{1-6}
{\tiny \yng(1,1,1,1)}&1&-1&1 &1&-1&\\
\cline{1-6}
{\small $\uparrow$ irrep}\\
\end{tabular}
\caption{Characters for each (irrep, class) 
pair of the permutation group $S_4$.}
\label{tab:perm-4}
\end{center}
\end{table}

The irreps of the unitary group $U(N)$ (and also 
the irreps of  $GL(N)$) are in one-to-one onto
correspondence with the Young graphs with at most $N$ rows.
We will specify the irreps of $U(N)$ by
$\eta^N$, where $\eta$ is an $N$-dimensional 
column vector of integers that specifies a Young graph. (The superscript 
$N$ serves to distinguish this from an irrep of a symmetric group).
In terms of tensors, the number $K$ of boxes  
of the Young graph corresponds to the number of 
tensor indices, $\{1, 2, \ldots, N\}$ corresponds to the range of the 
tensor indices, and the Young graph gives the symmetry 
properties of the tensor.

Weyl showed that the dimension of the irrep $\eta^N$  of $U(N)$
is given by:

\beq
\rmdim(\eta^N)= 
\frac{
\det((\eta+\delta)^{N-1}, (\eta+\delta)^{N-2}, \ldots, (\eta+\delta)^0)
}{\calF_{N-1}}
\;,
\label{eq:dim-eta}\eeq
where $\delta =(N-1, N-2, \ldots, 1, 0)^T$.

Let $\chi^{\eta^N}(A)$ represent the character of 
$A\in U(N)$ in the irrep $\eta^N$  of $U(N)$. 
Inspired by an identity due to Frobenius, Weyl 
derived the following  two expressions for 
$\chi^{\eta^N}(A)$. First,

\beq
\chi^{\eta^N}(A) =
\frac{
\det(\alpha^{\eta_0+N-1}, \alpha^{\eta_1+N-2}, \ldots, \alpha^{\eta_{N-1}})
}{
\det(\alpha^{N-1},\alpha^{N-2}, \ldots, \alpha^0)}
\;,
\label{eq:chi-ratio}\eeq
where $\alpha$ is the vector of eigenvalues of the 
$N\times N$ matrix $A$. Second,

\beq
\chi^{\eta^N}(A) =
\sum_{\calC(i)} \frac{|\;\calC(i)|}{K!}
\chi^{\eta}(\calC(i))
t_1^{i_1}
t_2^{i_2}
\cdots
t_K^{i_K}
\;,
\label{eq:chi-t-expan}\eeq
where $K$ is the number of boxes in the Young graph $\eta$, and
where the sum is over all classes 
$\calC(i)$ of the permutation group $S_K$,
$|\;\calC(i)|$ is the order of class $\calC(i)$,
$\chi^{\eta}(\calC(i))$ is the character of 
$\calC(i)$ in the irrep $\eta$  of $S_K$,
and where $t_1, t_2, \cdots$ are defined by

\beq
t_r = \tr(A^r) =\sum_{j=0}^{N-1} (\alpha_j)^r
\;,
\label{eq:trace-o-power}\eeq
where  $\{\alpha_j\}_j$ are the eigenvalues of $A$.

 Eq.(\ref{eq:dim-eta})
for $\rmdim(\eta^N)$ follows from Eq.(\ref{eq:chi-ratio})
and the fact that $\rmdim(\eta^N)=\chi^{\eta^N}(I)$,
where $I$ is the identity   matrix.
Let RHS1 denote the RHS of Eq.(\ref{eq:chi-ratio})
and RHS2 denote the RHS of Eq.(\ref{eq:chi-t-expan}).
Frobenius was the first to prove that RHS1=RHS2, but his proof,
which is discussed in Hamermesh\cite{Ham},
made no mention of $U(N)$. 
Weyl  gave a new proof\cite{Weyl} in which 
$U(N)$ was crucial. 

Note that in Eqs.(\ref{eq:chi-ratio})
 and (\ref{eq:chi-t-expan}),
$A$ is a unitary matrix. Hence, its eigenvalues are of 
the form $\alpha_j = e^{i\angle \alpha_j}$ for some real $\angle \alpha_j$.
However, the equation RHS1=RHS2 can be
analytically continued to complex $\alpha_j$ with $|\alpha_j| \neq 1$.
This is because both RHS1 and RHS2 are linear combinations of 
monomials of the form
$\alpha_0^{p_0}\alpha_1^{p_1}\ldots\alpha_{N-1}^{p_{N-1}}$,
where $p_0, p_1, \ldots, p_{N-1}$ are non-negative integers.
Since the $\angle \alpha_j$'s are arbitrary reals,  
the coefficient in RHS1
of any fixed monomial 
$\alpha_0^{p_0}\alpha_1^{p_1}\ldots\alpha_{N-1}^{p_{N-1}}$
 must equal
the coefficient in RHS2 of 
that same monomial.

 Table \ref{tab:char-u-n} was calculated using Eq.(\ref{eq:chi-t-expan}). 
A very similar table can be found in Ref.\cite{ItZu80}.
Table \ref{tab:dim-times-char} was derived from the information in 
Table \ref{tab:char-u-n}.

\begin{table}[h]
\begin{center}
\begin{tabular}{c|c|c|c}
$K$ &{\small irrep}& $\chi^{\eta^N}(A)$ & $\rmdim(\eta^N)=\chi^{\eta^N}(I)$\\
\hline
1 & {\tiny \yng(1)}& 
    $t_1$ & 
    $N$\\
\hline
2 & {\tiny \yng(2)}&
    $\frac{1}{2}t_1^2 + \frac{1}{2}t_2$ &
    $\frac{1}{2}N(N+1)$\\
\cline{2-4}
& {\tiny \yng(1,1)}&
    $\frac{1}{2}t_1^2 - \frac{1}{2}t_2$ &
    $\frac{1}{2}N(N-1)$\\
\hline
3 & {\tiny \yng(3)}& 
    $\frac{1}{6} t^3_1 +\frac{1}{2}t_1 t_2 +\frac{1}{3}t_3$&
    $\frac{1}{6}N(N+1)(N+2)$\\
\cline{2-4}
& {\tiny \yng(2,1)}& 
    $\frac{1}{3} t^3_1+ 0 t_1 t_2 -\frac{1}{3}t_3$&
    $\frac{1}{3}N(N+1)(N-1)$\\
\cline{2-4}
& {\tiny \yng(1,1,1)}& 
    $\frac{1}{6} t^3_1-\frac{1}{2} t_1 t_2 +\frac{1}{3}t_3$&
    $\frac{1}{6}N(N-1)(N-2)$\\
\hline
4 & {\tiny \yng(4)}& 
    $\frac{1}{24} t^4_1 +\frac{1}{4}t_1^2 t_2 
    +\frac{1}{8}t_2^2 + \frac{1}{3}t_1 t_3 + \frac{1}{4}t_4$&
    $\frac{1}{24}N(N+1)(N+2)(N+3)$\\
\cline{2-4}
& {\tiny \yng(3,1)}& 
    $\frac{1}{8} t^4_1 +\frac{1}{4}t_1^2 t_2
    -\frac{1}{8}t_2^2 + 0t_1 t_3 - \frac{1}{4}t_4$&
    $\frac{1}{8}N(N+1)(N+2)(N-1)$\\
\cline{2-4}
& {\tiny \yng(2,2)}& 
    $\frac{1}{12} t^4_1 +0t_1^2 t_2
    +\frac{1}{4}t_2^2 -\frac{1}{3}t_1 t_3 +0t_4$&
    $\frac{1}{12}N^2(N+1)(N-1)$\\
\cline{2-4}
& {\tiny \yng(2,1,1)}& 
    $\frac{1}{8} t^4_1 -\frac{1}{4}t_1^2 t_2
    -\frac{1}{8}t_2^2 +0t_1 t_3 +\frac{1}{4}t_4$&
    $\frac{1}{8}N(N+1)(N-1)(N-2)$\\
\cline{2-4}
& {\tiny \yng(1,1,1,1)}& 
    $\frac{1}{24} t^4_1 -\frac{1}{4}t_1^2 t_2 
    +\frac{1}{8}t_2^2 + \frac{1}{3}t_1 t_3 - \frac{1}{4}t_4$&
    $\frac{1}{24}N(N-1)(N-2)(N-3)$\\
\end{tabular}
\caption{$K$ is the number of boxes in the Young graph. 
The irreps of $U(N)$ and $GL(N)$ are in 1-1 correspondence with
the Young graphs with at most $N$ rows. 
$\chi^{\eta^N}(A)$ is the character of $A\in U(N)$ in 
the irrep  $\eta^N$ of $U(N)$.
$\rmdim(\eta^N)$ is the dimension of  irrep $\eta^N$.}
\label{tab:char-u-n}
\end{center}
\end{table}

\begin{table}[h]
\begin{center}
\begin{tabular}{c|l}
$K$ & $\sum_{{\rm all\; irreps}\;\eta^N\;{\rm with}\;  K\; {\rm boxes}}
\rmdim(\eta^N) \chi^{\eta^N}(A)$\\
\hline
0&1 (convenient definition)\\
\hline
1&$ N t_1$\\
\hline
2 & $\frac{N}{2} (N t_1^2 + t_2)$\\
\hline
3 & $\frac{N}{6} (N^2 t_1^3 + 3 N t_1 t_2  + 2 t_3)$\\
\hline
4 & $\frac{N}{24}( N^3 t_1^4 + 6 N^2 t_2 t_1^2 + 8 N t_3 t_1 + 3 N t_2^2 + 6 t_4)$\\
\hline
\end{tabular}
\caption{This table was derived 
using the information in Table \ref{tab:char-u-n}.}
\label{tab:dim-times-char}
\end{center}
\end{table}

\section{Ensemble of Quantum Density Matrices}
In this section, we will generalize the 
simplex moments integral Eqs.(\ref{eq:sim-int})
to
the quantum realm. To go from classical to quantum physics,
we will replace
probability distributions by quantum density matrices.
In the quantum case, we will need to do integrals over a
manifold of matrices. Such integrals are used in 
several fields of mathematical physics. They are crucial to 
 the field of Random Matrix Theory \cite{MeBook}.

Let
\beq
U = \frac{1}{\sqrt{2}}
\left(
\begin{array}{cc}
1 & i \\
1 & -i 
\end{array}
\right)
\;.
\eeq
It is easy to check that $U$ is unitary.
For any complex number $z$,
let $z_\Re$ (ditto, $z_\Im$) represent its real (ditto, imaginary) part.
Note that

\beq
\left(
\begin{array}{c}
z\\ z^*
\end{array}
\right)
=
U
\left(
\begin{array}{c}
z_\Re\sqrt{2}\\ z_\Im\sqrt{2}
\end{array}
\right)
\;.
\eeq
This transformation rule 
motivates us
to generalize the volume element $dx$
in real space to a volume element $d^2 z$
in complex space according to:

\beq
dx \rarrow d^2 z = 2 dz_\Re dz_\Im
\;.
\eeq
We also generalize the Dirac delta function $\delta(x)$
in real space to a Dirac delta function $\delta^2(z)$
in complex space:

\beq
\delta(x)\rarrow \delta^2(z) =\frac{1}{2}\delta(z_\Re)\delta(z_\Im)
\;.
\eeq
Thus,

\beq
\int^\infty_{-\infty} dx\;\delta(x) = 1 \rarrow
\int d^2 z \; \delta^2(z) = 1
\;.
\eeq

Next we want to give a convenient 
definition of a volume element $\fatd \rho$ for the
manifold of all Hermitian $N\times N$ matrices $\rho$.
We define 

\beq
\fatd \rho =\prodset{d\rho_{xx}}{x}
\prodset{d^2\rho_{xx'}}{x,x' : \;x<x'}
\;.
\eeq
This  definition of $\fatd \rho$ prompts us to define:

\beq
\delta(\rho) =\prodset{\delta(\rho_{xx})}{x}
\prodset{\delta^2(\rho_{xx'})}{x,x' : \;x<x'}
\;,
\eeq
so that

\beq
\int \fatd\rho \;\delta(\rho) = 1
\;.
\eeq

The manifold of real points $x$ 
with volume element $dx$
corresponds to the manifold of
Hermitian $N\times N$ matrices $\rho$ with volume
element $\fatd\rho$. But what is the matrix counterpart of
the manifold of complex points $z$ with 
volume element $d^2 z$? A natural candidate for this is the manifold of all
complex $N\times N$ matrices $A$. We define
its volume element by:

\beq
\fatd^2 A=
\prodset{d^2A_{x x'}}{x, x'}
\;.
\eeq
Note that $\fatd^2 A$ is a product of
twice as many $dx$-like
real-space volume elements as $\fatd^2 \rho$. 
Finally, we define

\beq
\delta^2(A) =
\prodset{\delta^2(A_{x x'})}{x, x'}
\;
\eeq
so that

\beq
\int \fatd^2 A \;\delta^2(A) = 1
\;.
\eeq

Given two  Hermitian $N\times N$ matrices $\rho$ and
$\omega$, one has 
\beq
\omega\cdot \rho = 
\sum_x \omega_{xx}\rho_{xx} 
+ 2 \sum_{x<y}\left[ 
(\omega_{xy})_\Re (\rho_{xy})_\Re - 
(\omega_{xy})_\Im (\rho_{xy})_\Im 
\right]
\;.
\eeq
Combining  the last equation and the identity 
$\delta(x) = \int_{-\infty}^\infty \frac{dk}{2\pi}e^{ikx}$
for real $x$, yields

\beq
 \int \fatd\omega \;\exp(i \omega\cdot\rho)
= (2\pi)^{N^2}\delta(\rho)
\;.
\eeq

Let $X$  be a Hermitian $N\times N$ matrix
with eigenvalues $\{\chi_j\}_j$. 
We need to consider a real valued function $f(X)$ that
is invariant under unitary transformations of its  argument $X$; that is,
$f(UXU^\dagger)= f(X)$ for any unitary $N\times N$ matrix $U$.
$f(X)$ depends only on the eigenvalues of $X$, so $f(X) = F(\chi)$
for some function $F: \mathbf {R}^N\rarrow \mathbf {R}$.
Note that not all functions from 
$\mathbf {R}^N$ to $\mathbf {R}$ qualify for the job of $F$.
$F(\chi)$ must also depend on $\chi$
in a  symmetrical way: since all $N\times N$ permutation matrices belong
to $U(N)$, $F(\chi)$ must be invariant 
under permutations of its arguments $\{\chi_j\}_j$.
Henceforth, 
we will indulge in a convenient abuse
of notation by replacing the symbol $F$ by $f$ 
so that $f(X) = f(\chi)$.

Let $X$ and $A$ be Hermitian $N\times N$ matrices
with eigenvalues $\{\chi_j\}_j$ and $\{\alpha_j\}_j$,
respectively. Let $f(X)$ be a real valued function of $X$ that
is invariant under unitary transformations of its argument $X$.
Finally, let $\xi$ be an arbitrary complex number.
The following two integral formulas are well known:
First(see Ref.\cite{MeBook}),

\beq
\int \fatd X \;f(X) = 
\frac{(2\pi)^{L_N}}{\calF_N}
\prodset{\int_{-\infty}^\infty d\chi_j}{j}
\Delta^2(\chi) f(\chi)
\;,
\label{eq:int-f}\eeq
and, second(see Refs.\cite{ItZu80}
\and \cite{Me81}),

\beq
\int \fatd X \; e^{\xi \tr(AX)} f(X) =
\left(\frac{2\pi}{\xi}\right)^{L_N}
\prodset{\int_{-\infty}^\infty d\chi_j}{j}
\frac{\Delta(\chi)}{\Delta(\alpha)}
e^{\xi (\chi \cdot \;\alpha)} f(\chi)
\;.
\label{eq:int-f-exp}\eeq
In the spirit of non-rigorous, applied mathematics,
I will not specify 
precise sufficiency conditions on
$A$, $f(\cdot)$ and
$\xi$ under which 
Eqs.(\ref{eq:int-f}) and (\ref{eq:int-f-exp})
are valid. I leave it to more competent 
pure mathematicians to 
figure this out. 

Proofs of Eqs.(\ref{eq:int-f}) 
and (\ref{eq:int-f-exp}) 
 are easily found in the literature
so we won't present them here. 
Proving Eq.(\ref{eq:int-f-exp}) 
involves solving a partial differential equation
(an initial value problem for a diffusion equation).
 
Of course, Eq.(\ref{eq:int-f}) follows from
Eq.(\ref{eq:int-f-exp}) when $\xi\rarrow 0$.
To check this quickly,
assume (without loss of generality) that
the eigenvalues $\alpha_j$ are very small and widely separated:

\beq
0<\alpha_0\ll \alpha_1 \ll \cdots\ll \alpha_{N-1}\ll 1
\;.
\eeq
Then, keep the largest term in the expansion of $\Delta(\alpha)$:

\beq
\Delta(\alpha)\approx \alpha_0 \alpha_1 \ldots \alpha_{N-1}
\;.
\eeq
Furthermore, keep  only the largest term in
the Taylor expansion of $\exp(\xi\chi\cdot \alpha)$ 
and in the multinomial expansion of
$(\chi\cdot\alpha)^{L_N}$:

\beq
\exp(\xi\chi\cdot \alpha) 
\approx \frac{(\xi\chi\cdot\alpha)^{L_N}}{L_N!}
\approx \frac{\xi^{L_N}}{L_N!}
\frac{L_N!}{\calF_{N-1}} 
(\chi_0\chi_1\cdots\chi_{N-1})
(\alpha_0\alpha_1\cdots\alpha_{N-1})
\;.
\eeq
Use these approximations on the integrand on the RHS of 
Eq.(\ref{eq:int-f-exp}). Also replace  the $\chi$
dependent part of the integrand, that is,
$\Delta(\chi)\chi_0\chi_1\cdots\chi_{N-1}$,
by its totally symmetric part $\Delta^2(\chi)/N!$.
This converts the RHS of Eq.(\ref{eq:int-f-exp})
into the RHS of Eq.(\ref{eq:int-f}) in the limit $\xi\rarrow 0$.

We are almost ready to present our generalization of
the simplex moments integral. But first we need to prove 
two lemmas.

\begin{lemma}
Suppose $\beta_0, \beta_1, \ldots, \beta_{N-1}$ are non-negative integers. Then
\beq
\epsilon_{k_0 k_1 \cdots k_{N-1}}
\prodset{(k_j + \beta_j)!}{j}
=
\prodset{\beta_j\;!}{j} \Delta(\beta)
\;
\label{eq:det-lemma}\eeq
\end{lemma}
proof:

Note that the LHS of Eq.(\ref{eq:det-lemma}) can be
expressed as $\det(M)$, where the matrix $M$ has entries
$M_{i,j} = (i + \beta_j)!$. 
Eq.(\ref{eq:det-lemma})
can be  proven easily using mathematical induction
 and simple properties of determinants.
QED

\begin{lemma}
Suppose $\beta_0, \beta_1, \ldots, \beta_{N-1}$ are non-negative integers. Then
\beq
\prodset{\int_{0}^\infty dx_j}{j}
\delta(\sum_j x_j -1)
\Delta(x) x_0^{\beta_0} x_1^{\beta_1} \ldots x_{N-1}^{\beta_{N-1}}
=
\frac{ \prodset{\beta_j !}{j} \Delta(\beta)}
{(\sum_j \beta_j + L_N + N -1)!}
\;
\label{eq:int-lemma}\eeq
\end{lemma}
proof:

Replace $\Delta(x)$ by

\beq
\Delta(x) = \epsilon_{k_0 k_1 \cdots k_{N-1}}
x_0^{k_0} x_1^{k_1} \ldots x_{N-1}^{k_{N-1}}
\;.
\eeq
in the LHS of Eq.(\ref{eq:int-lemma}). Then use the
simplex moments integral  Eqs.(\ref{eq:sim-int})  to get
\beq
LHS =  \frac{\epsilon_{k_0 k_1 \cdots k_{N-1}}
\prodset{(k_j + \beta_j)!}{j}
}{\nu}
\;,
\eeq
where

\beq
\nu = \sum_j (\beta_j + k_j) + N -1 = \sum_j \beta_j + L_N + N-1
\;.
\eeq
Finally, apply Eq.(\ref{eq:det-lemma}). 
QED

The volume of a simplex 
($\nu_b=0\;\forall b$ in Eq.(\ref{eq:sim-int}) )
is the simplest 
 case of the 
simplex moments integral. To warm up,
we first present the quantum version of this simplest case:

\begin{claim}
Suppose $X\in Herm(N)$.
Then

\beq
\int \fatd X \;\theta(X\geq 0)\delta( \tr X -1)= 
\frac{(2\pi)^{L_N}\calF_{N-1}}
{(N^2-1)!}
\;.
\label{eq:quant-vol-int}
\eeq
\end{claim}
proof:

Let LHS (ditto, RHS)
stand for the left (ditto, right) hand side of Eq.(\ref{eq:quant-vol-int}). Then

\begin{subequations} 
\beqa 
LHS &=& \frac{(2\pi)^{L_N}}{\calF_N} 
\prodset{\int_0^\infty d\chi_j}{j} 
\Delta^2(\chi)
 \delta(\sum_j\chi_j -1)\\
&=& \frac{(2\pi)^{L_N}}{\calF_N} 
N! 
\prodset{\int_0^\infty d\chi_j}{j} 
\Delta(\chi) \chi_0^0 \chi_1^1\ldots\chi^{N-1}_{N-1}
\delta(\sum_j\chi_j -1)\\
&=& \frac{(2\pi)^{L_N}\Delta(0, 1, 2, \ldots, N-1)  }{(2L_N +N-1)!}\\
&=& RHS
\;.
\eeqa 
\end{subequations}
In going from line $a$ to line $b$, we replaced
one of the $\Delta(\chi)$
by $N! \chi_0^0 \chi_1^1\ldots\chi^{N-1}_{N-1}$; this was valid 
 because the rest of the integrand was 
totally anti-symmetric under permutations of the $\{\chi_j\}_j$.
To go from line $b$ to line $c$,
we applied Eq.(\ref{eq:int-lemma}). To go from line $c$ to line $d$, we used 
$\Delta(0, 1, 2, \ldots, N-1) = \calF_{N-1}$.
QED

Finally, we are ready to 
present the main result of this paper,
a generalization of the simplex moments
integral to quantum mechanics.
Actually, we will give a moment generating
function and calculate moments from that.

\begin{claim}
Suppose $X, A\in Herm(N)$, and
$\xi$ is a complex number. 
If both sides of the following formula exist, then

\beq
\int \fatd X e^{\xi tr(AX)} \theta(X\geq 0) \delta( \tr X -1)= 
\sum_{K=0}^\infty  \xi^K
\frac{ (2\pi)^{L_N}\calF_{N-1}}
{(K+N^2-1)!}
\sum_{\eta^N : K\;{\rm boxes}}
\rmdim(\eta^N) \chi^{\eta^N}(A)
\;,
\label{eq:quant-mom-int}\eeq
where we define $\rmdim(\eta^N) \chi^{\eta^N}(A)
=1$ for $K=0$.
\end{claim}
proof:

We will prove this claim when $\xi=1$. The more general case can be obtained
from this by scaling $A$ (i.e., replacing $A$ by $\xi A$, where $\xi$ is
real, and then analytically continuing $\xi$ to complex values.)
Let LHS (ditto, RHS)
stand for the left (ditto, right) hand side of Eq.(\ref{eq:quant-mom-int}).

Applying Eq.(\ref{eq:int-f-exp}) yields

\beq
LHS = (2\pi)^{L_N}
\prodset{\int_{-\infty}^\infty d\chi_j}{j}
\frac{\Delta(\chi)}{\Delta(\alpha)}
e^{ \chi \cdot \;\alpha} 
\prodset{\theta(\chi_j\geq 0)}{j}
\delta(\sum_j \chi_j -1)
\;.
\label{eq:after-int-f-exp}
\eeq

Recall the multinomial expansion:

\beq
(x_1 + x_2 + \cdots + x_N)^K = 
\sum_{\vec{k}} 
\delta(\sum_{r=1}^N k_r, K)
\frac{K!}{\prodset{k_r!}{r}} 
x_1^{k_1}x_2^{k_2}\cdots x_N^{k_N}
\;.
\eeq
Using the Taylor expansion of $\exp(\cdot)$
and the multinomial expansion, we get

\begin{subequations}
 \beqa
\exp(\chi\cdot\alpha) &=& \sum_{K'=0}^{\infty} 
\frac{(\chi\cdot\alpha)^{K'}}{K'!}\\
&=&\sum_{K'=0}^{\infty}
\sum_{\vec{k}}
\delta(\sum_r k_r, K') 
\prodset{ \frac{(\chi_r \alpha_r)^{k_r}}{k_r!}}{r}
\;.
\eeqa 
\end{subequations}
Now apply to Eq.(\ref{eq:after-int-f-exp})
this expansion of $\exp(\chi\cdot\alpha) $ and Eq.(\ref{eq:int-lemma}):

\beq
LHS = 
(2\pi)^{L_N} \sum_{K'=0}^{\infty}\sum_{\vec{k}}
\delta(\sum_r k_r, K')
\frac{1}{\Delta(\alpha)}
\prodset{\alpha_r^{k_r}}{r}
\frac{\Delta(\vec{k})}{(\sum_r k_r + L_N + N-1)!}
\;.
\eeq

The function being summed over $\vec{k}$
is a product of a totally anti-symmetric function of 
$\vec{k}$ times
 $\prodset{\alpha_r^{k_r}}{r}$.
Hence we may replace $\prodset{\alpha_r^{k_r}}{r}$
by its totally anti-symmetric part:

\beq
\prodset{\alpha_r^{k_r}}{r}
\rarrow
\frac{1}{N!} \det(\alpha^{k_0}, \alpha^{k_1},\ldots, \alpha^{k_{N-1}})
\;.
\eeq
After doing this we will have a sum over 
$\vec{k}$ of a totally symmetric function of 
$\vec{k}$ so we can replace:

\beq
\sum_{\vec{k}}\rarrow 
N! \sum_{\vec{k}} \theta(0\leq k_0<k_1<\cdots<k_{N-1})
\;.
\eeq
(Terms with $k_i=k_{i+1}$  vanish.) 
We can also change the lower limit of the
$K'$ sum from $K'=0$ to $K'=L_N$,
because terms with $K'<L_N$ do not contribute.
Let $K = K' - L_N$. Let us change variables from $K'$ to $K$.
All these changes yield:

\beqa
LHS &= &
(2\pi)^{L_N}
\sum_{K=0}^{\infty}
\sum_{\vec{k}} \nonumber\\
&&
\theta(0\leq k_0<k_1<\cdots<k_{N-1})
\delta(\sum_r k_r, K + L_N) \nonumber\\
&&
\frac{\Delta(\vec{k})}{(K + N^2 -1)!}
\frac{
\det(\alpha^{k_0},\alpha^{k_1}, \ldots, \alpha^{k_{N-1}}) 
}{
\det(\alpha^{0},\alpha^{1}, \ldots, \alpha^{N-1}) 
}
\;.\label{eq:pre-eta-to-k-line}
\eeqa 

Next consider the following change of variables:

\beq
\begin{array}{l}
\eta_{N-1} = k_0\\
\eta_{N-2} = k_1 -1\\
\;\;\;\vdots\\
\eta_1 = k_{N-2} - (N-2)\\
\eta_0 = k_{N-1} - (N-1)
\end{array}
\;.
\label{eq:eta-to-k}\eeq
Note that $\eta_0- \eta_1 = k_{N-1}- k_{N-2}-1\geq 0$,
$\eta_1- \eta_2 = k_{N-2}- k_{N-3}-1\geq 0$, etc.
Also, $\sum_j \eta_j =K' - L_N = K$.
Hence, $\eta$ specifies a Young graph with $K$
boxes and at most $N$ rows.

Define a column vector $\delta = (N-1,N-2,\cdots,1,0)^T$.
Let $R$ be the $N\times N$  matrix that has ones on the non-principal
diagonal and zeros everywhere else. For example, for $N=2$,
$R=
\left(
\begin{array}{ccc}
0&1\\
1&0\\
\end{array}
\right)$.
For any $N$ dimensional column vector
$\eta$, define its $R$ transform $\eta^R$ by $\eta^R = R\eta$.
$R$ just reverses the entries of $\eta$.
Eqs.(\ref{eq:eta-to-k})
can now be stated succinctly as $\eta = k^R - \delta$.
Note that
$R^2=1$.
Thus, $\det(RAR)=\det(A)$, for any square matrix 
$A$ of the same dimension as $R$. The right and left $R$'s reverse
the order of the columns and of the rows of $A$, but this
does not change
the value of $\det(A)$.
For example, suppose $\det(A)$ is the Vandermonde determinant $\Delta(v)$ of a vector
$v$. Assume $v$ is 3 dimensional for concreteness. Then
$\det(v^0, v^1, v^2) = \det((v^R)^2, (v^R)^1, (v^R)^0)$.

Changing the summation variable from $\vec{k}$ to $\eta$
in Eq.(\ref{eq:pre-eta-to-k-line}) yields:

\beq
LHS= 
\frac{(2\pi)^{L_N}\calF_{N-1} }{(N^2-1)!}
+
(2\pi)^{L_N}
\sum_{K=1}^{\infty}
\sum_{\eta^N}
\frac{\Delta(\eta^R +\delta^R)}{(K + N^2 -1)!} \gamma
\;, \label{eq: reversed-del-gamma}
\eeq
where $\gamma$ is defined by

\beq
\gamma = 
\frac{
\det(\alpha^{\eta_{N-1}},\alpha^{\eta_{N-2}+1}, \ldots, \alpha^{\eta_0+N-1}) 
}{
\det(\alpha^{0},\alpha^{1}, \ldots, \alpha^{N-1}) 
}
\;.
\eeq

Reversing the order of the columns of both the denominator and numerator
determinants of $\gamma$ yields:

\beq
\gamma=
\frac{
\det(\alpha^{\eta_0+N-1},\alpha^{\eta_1+N-2}, \ldots, \alpha^{\eta_{N-1}}) 
}{
\det(\alpha^{N-1},\alpha^{N-2}, \ldots, \alpha^{0}) 
}
=
\chi^{\eta^N}(A)
\;.
\label{eq:new-gamma}
\eeq
Reversing columns and rows in $\Delta(\eta^R + \delta^R)$ leads
to Weyl's formula for the dimension of $\eta^N$.

\beq
\Delta(\eta^R + \delta^R) =
\det[ (\eta+\delta)^{N-1}, (\eta+\delta)^{N-2}, \ldots, (\eta+\delta)^0]
= \rmdim(\eta^N) \calF_{N-1}
\;.
\label{eq:new-delta}
\eeq
Applying Eqs.(\ref{eq:new-gamma}) and
(\ref{eq:new-delta})  to Eq.(\ref{eq: reversed-del-gamma}) finally yields
LHS= RHS.
QED

Eq.(\ref{eq:quant-mom-int}) gives a generating function that 
can be used to calculate moments over a uniform ensemble of density matrices.
For example, we can calculate the mean value of $X_{i_1,j_1} X_{i_2, j_2}$.
Such moments will have indices attached because $X$ is a matrix.
To avoid having free indices, we will bind them to 
constant matrices.
So instead of calculating the mean value of $X_{i_1,j_1} X_{i_2, j_2}$,
we will calculate the mean value of $(C_1\cdot X)( C_2\cdot X)$,
where $C_1$ and $C_2$ are constant  matrices.

For any $N\times N$ matrices $A$ and $C$, consider how the operator
$C\cdot \pder{}{A}$ acts on a power $A^n$
for some integer $n$. $C\cdot \pder{}{A}$ 
replaces one $A$ at a time by a $C$:

\beq
C\cdot \pder{}{A} A^n =
\sum_{i=0}^{n-1} A^i C A^{n-1-i}
\;.
\eeq
For example, if we define the operator $\Omega$ by

\beq
\Omega =  
\prodset{C_j \cdot \pder{}{A}}{1\leq j\leq 4}
\;,
\eeq
then:

\begin{subequations}\label{eq:omega-a-4}
\beq
\Omega t_1^4 = \sum_P \tr(C_1) \tr(C_2) \tr(C_3) \tr(C_4)
\;,
\eeq

\beq
\Omega t_2 t_1^2= \sum_P \tr(C_1 C_2) \tr(C_3) \tr(C_4)
\;,
\eeq

\beq
\Omega t_2^2 = \sum_P \tr(C_1 C_2) \tr(C_3 C_4)
\;,
\eeq

\beq
\Omega t_3 t_1= \sum_P \tr(C_1 C_2 C_3) \tr(C_4)
\;,
\eeq

\beq
\Omega t_4 = \sum_P \tr(C_1 C_2 C_3 C_4)
\;.
\eeq
\end{subequations}
In Eqs.(\ref{eq:omega-a-4}), $t_r = \tr(A^r)$ as before,
and the sums run over all permutations $P$ on 4 letters.
$P$ acts on the subscripts $\{1,2,3,4\}$.

Define

\beq
Z(A) = \int \fatd X \; e^{\tr(AX)} \theta(X\geq 0) \delta(\tr X -1)
\;,
\eeq

\beq
I(C_1) = \int \fatd X \; (C_1\cdot X)\theta(X\geq 0) \delta(\tr X -1)
\;,
\eeq

\beq
I(C_1, C_2) = \int \fatd X \; (C_1\cdot X)(C_2\cdot X) \theta(X\geq 0) \delta(\tr X -1)
\;.
\eeq
Then, by virtue of Table \ref{tab:dim-times-char}
and Eq.(\ref{eq:quant-mom-int}), one has

\begin{subequations} 
\beqa
I(C_1) &=& \lim_{A\rarrow 0}
\left(C_1\cdot\pder{}{A}\right)
Z(A)\\
&=& \frac{(2\pi)^{L_N} \calF_{N-1}}
{(N^2)!} 
N \tr(C_1)
\;,
\eeqa 
\end{subequations}
and

\begin{subequations} 
\beqa
I(C_1, C_2) &=& \lim_{A\rarrow 0}
\left(C_1\cdot\pder{}{A}\right)
\left(C_2\cdot\pder{}{A}\right)
Z(A)\\
&=& \frac{(2\pi)^{L_N} \calF_{N-1}}
{(N^2+1)!}
 N[ N\tr(C_1)\tr(C_2) + \tr(C_1 C_2)]
\;.
\eeqa 
\end{subequations}

\end{document}